\documentclass[12pt,a4paper]{article}
\usepackage{amsmath,amssymb,amscd}
\usepackage{graphicx,epsfig}

\begin{document}

\begin{center}
{\bf DECOHERENCE AND ENERGY LOSS IN QCD CASCADES IN NUCLEAR COLLISIONS}
\end{center}

\bigskip

\begin{center}

{\bf A.~Leonidov$^{(a,b)}$, V.~Nechitailo$^{(a,b)\,}$}

\bigskip

\bigskip

(a) {\it Theoretical Physics Department, P.N.~Lebedev Physics Institute, \\
Moscow, Russia}

(b) {\it Institute of Theoretical and Experimental Physics, Moscow, Russia}

\end{center}

\bigskip

\bigskip

\begin{center}
{\bf Abstract}
\end{center}

The medium modifications in the properties of QCD cascades are considered. In
particular, the changes in the intrajet rapidity distributions due to
medium-induced decoherence, collisional losses of cascade gluons and those of
final prehadrons are analyzed.  

\newpage

\section{Introduction}

One of the most exciting experimental discoveries at RHIC are the dramatic
changes in the patterns of energy flow in the events with large transverse
energy. The simplest and most discussed related effect is the so-called jet
quenching, the attenuation of hadrons with moderated and large transverse
momenta, see the recent experimental review \cite{E09} and the theoretical
reviews \cite{BSZ00,GVWZ03,KW03,SS07,W09,M10} devoted to this problem.
Generically the events with large transverse energies are jetty, i.e. are
characterized by collimated bunches of hadrons.  A description of intrajet
properties of high energy jets in usual QCD vacuum \cite{DKMT91} belongs to the
most successful applications of perturbative QCD.  The detailed picture of parton
cascade dynamics including fine details such as a destructive quantum
interference leading to an angular ordering in sequential gluon decays was
worked out and shown to successfully explain experimental data on intrajet
spectra and multiplicity distributions. This explains a great interest in studying
the modifications of intrajet properties of QCD jets produced in high energy
nuclear collisions in which QCD cascades develop in high density environment.
Experimental studies of this problem have been performed at RHIC \cite{JSTAR}
and are planned at LHC \cite{JLHC}.

On the theoretical side, in order to study the medium-induced modifications, one
needs to work out the corresponding generalization of the formalism of the in-vacuum QCD. The
presence of dense medium could influence the evolution of a cascade in two ways.
First, the presence of the medium
facilitates cascading by creating additional phase space for gluon emission. 
Second, there could be a non-radiative energy loss (gain) due to elastic
interactions of the projectile with the modes of the medium.  To
account for interactions of the perturbative degrees of freedom in the evolving
jet with the nonperturbative soft modes and/or medium one has to introduce an
explicit mechanism of interaction between the partonic cascade and the
environment.

The most detailed studies of in-vacuum QCD cascades are performed with the help of the Monte-Carlo
simulations. The two most popular versions of these MC generators are PYTHIA
\cite{PYTHIA} and HERWIG \cite{HERWIG}. Both of them implement a key property of
in-vacuum QCD jets, the angular ordering, that in the modern versions is
directly built in, see a detailed discussion in \cite{QHERWIG} and in the
Appendix of the present paper. The first PYTHIA-based MC program including the
effects of medium-induced radiative energy loss was PYQUEN \cite{PYQUEN}.
Recently there appeared several MC generators treating the medium-induced
effects in the QCD cascades, in particular the PYTHIA-based JEWEL \cite{JEWEL}
and Q-PYTHIA \cite{QPYTHIA} and the HERWIG-based "Q-HERWIG" \cite{QHERWIG}. In
the present paper we have used the object-oriented code realizing the 'global'
'constrained' mass-ordered version of PYTHIA implemented in JEWEL \cite{JEWEL}.
All of the above-listed MC generators take into account the medium-induced
radiative energy loss that is believed to be a major source of the energy loss; 
JEWEL, in addition, includes the effects of rescattering
of cascade particles on those in the medium.  The generic effect due to the
medium-related interactions is the softening of intrajet particle spectra
accompanied by the transverse broadening of jets. 

Let us note that for the case of non-radiative energy loss a corresponding
modification of Altarelli-Parisi equations, constructed in analogy with a theory
of ionization losses in electromagnetic cascades \cite{coscad}, was considered
in \cite{pionizationNAO,pionizationAO} where spectra and multiplicity
distributions of QCD jets in a dissipative medium were studied.

The focus of the present paper is on the analysis of some medium-induced
modifications in the evolution of QCD cascades related to the fine structure 
of the spatiotemporal pattern of the in-medium QCD cascade, in particular on the
substantial softening of the intrajet rapidity distribution. 

In the paragraph \ref{model} we describe the basic features of the model of
in-medium QCD cascade used in our simulations. 

In the paragraph \ref{moc} we give the detailed description of the simulation of
the mass-ordered QCD cascade.  

In the paragraph \ref{dc} we discuss the substantial softening of the intrajet
rapidity distributions due to decoherence effects leading to the disappearance
of the angular ordering. 

In the paragraph \ref{ic} we consider the effects of the collisional energy loss as related
to the spatiotemporal pattern of the in-medium QCD cascade.

We conclude the paper with the short summary of the results obtained and of the problems
to be addressed in the near future. 

\section{Decoherence and energy loss in in-medium QCD
cascade}\label{decoherence}

In the present paper we will concentrate on discussing the properties of the
timelike in-medium QCD cascades. The physical origin of the cascading process is
the intense gluon radiation of a high energy parton created in a hard
subprocess. Operationally one can think of this radiation cascade as of
equipping the initial parton with energy $E_0$ with some invariant mass $Q_0^2$
that is shaken off through decays into partons with lower invariant masses until
reaching some non-perturbative scale $Q_h/2$ at which preconfined colorless
clusters are formed.

When considering the properties of the in-medium QCD cascades one has to deal
with several mechanisms modifying the in-vacuum intrajet
characteristics. These are, in particular, the loss of quantum coherence and the
resulting loss of angular ordering, collisional and
radiative energy losses. In the present paper we shall focus on studying the
impact of decoherence and collisional energy loss of partons and final prehadrons.
The main motivation is to analyze the dependence of observable quantities such as
intrajet particle distributions, jet energy and multiplicity on the spatiotemporal 
pattern of the cascade, in particular on the distance
from the origin of the cascade to the border of the hot dense fireball or,
equivalently, the time interval during which the development of the cascade
takes place inside the medium.  To get a feeling for the corresponding physics,
let us consider, for example, a gluon with the energy $E_0=100$ GeV
equipped with a typical initial invariant mass $Q_0=10$ GeV. This gluon 
will on the average decay at $\tau_0 \sim E_0/Q_0^2=1\;{\rm GeV}^{-1}\simeq 0.2$ fm, so that if
the vertex in which the initial hard gluon was generated was sufficiently far
(several fermies) from the surface of hot fireball, a significant part of
cascade vertices will be generated inside the medium.

\subsection{The model}\label{model}

\begin{figure}[h]
 \centering
 \includegraphics[width=0.75\textwidth]{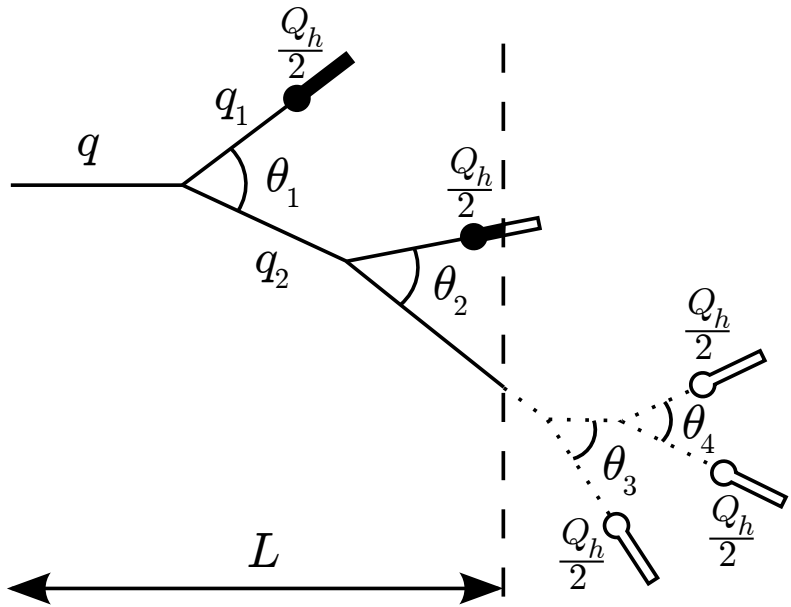}
 \caption{A sketch of the in-medium QCD cascade.}
 \label{fcascade}
\end{figure}

The key elements of the model developed in the paper are illustrated in Fig.~\ref{fcascade}.

The area to the left of the vertical dashed line is the region filled with dense
parton matter where modifications of cascade properties take place. To the right
of this line is the region of usual vacuum in which one has usual unmodified QCD
cascade. The cascade is assumed to be initiated at the distance $L$ from the
medium-vacuum interface surface.

The considered modifications of cascade properties inside the medium are related
to several mechanisms: 
decoherence due to color randomization, collisional losses of cascading gluons
and those of final prehadrons.

\subsubsection{Space-time pattern.}\label{stp}

To determine whether a given line is still inside the medium or has already left
it, one needs to set up a clock counting time along the cascade. The only
possibility of associating a spatiotemporal pattern with the cascade is to use
the lifetime $\tau$ of a virtual parton which, for a parton with the energy $E$
and virtuality $Q^2$ that has been created in the decay of its parent parton
with the virtuality $Q^2_{\rm par}$, reads
\begin{equation}\label{formtime}
\tau = E \left(\frac{1}{Q^2}-\frac{1}{Q^2_{\rm par}} \right)
\end{equation}
The lifetime of the initial parton is taken to be $\tau_{\rm
in}=E_{\rm in}/Q^2_{\rm in}$. For example, the time $\tau_h$ at
which the parton with the momentum $q_1$ in Fig.~\ref{fcascade} produced the
final prehadron is
$$
\tau_h = E_0 \frac{1}{Q_0^2} + E_1 \left(
\frac{4}{Q^2_h}-\frac{1}{Q_0^2}\right) = \frac{E_0}{Q_0^2} \left[ 1+z
\left( \frac{4Q_0^2}{Q^2_h}-1\right) \right]
$$
where $z\equiv E_1/E_0$.
For the parton under discussion it happened that $\tau_h < L$ so
that both its formation at time $E_0/Q^2_0$ and its decay into
final prehadron took place inside the medium.

\subsubsection{Decoherence due to color randomization}\label{cdl}

For decays taking place inside the medium such as those of gluons with the
momenta $q$ and $q_1$ the angular ordering is absent. This means that,
e.g., the magnitude of the angle $\theta_2$ is not restricted by that of the
angle $\theta_1$. The corresponding effects will be discussed in the
paragraph \ref{dc}.

\subsubsection{Collisional losses of cascading gluons}\label{egl}

The gluons propagating in the medium (solid lines) experience, before converting
to final prehadrons (black circles) collisional energy loss. The
corresponding effects will be considered in the paragraph \ref{ic}.

\subsubsection{Collisional losses of final prehadrons}\label{epl}

The final prehadrons formed inside the medium, such as, e.g., that formed by the
gluon initially having the momentum $q_1$, experience the medium-induced collisional
loss and, if having insufficient energy, do not leave the medium. The effects
related to the prehadron collisional energy loss are discussed in the paragraph
\ref{ic}.

\subsection{Mass-ordered QCD cascade}\label{moc}

In this paragraph we describe the procedure of generating a mass-ordered gluon
cascade we employ.

The cascade evolution goes through a sequence of decays $q \Rightarrow q_1+q_2$,
where $q=(E,\vec{q})$ is a four-momentum of the parent gluon and
$q_{1(2)}$ are those of the daughter ones. We study the timelike evolution in
which cascading reshuffles the initial jet virtuality in such a way that
at each decay one has $q^2 >q_{1}^2+q_{2}^2$, see below. The cascading process
at a given branch stops when a virtuality of the last parton reaches a
threshold level $Q_h^2$, 
i.e. no decay into two partons with minimal virtuality
$Q^2_h/4$ can occur\footnote{By default the virtuality of such a last parton is set to $Q^2_h/4$.}. The key property of the in-vacuum QCD cascades,
effective angular ordering of subsequent decays due to color coherence, is not
automatically taken into account in the mass-ordered evolution and has
to be enforced explicitly by accepting only those generated decays for which the
condition of angular ordering does hold.

As the angular ordering is checked only a posteriori, the restrictions imposed
on initially generated decays are purely kinematical so that
for the decay $q \to q_1 + q_2$ the splitting variable $z=E_1/E \equiv 1-E_2/E$
should lie in the interval
\begin{equation}\label{splitint0}
{\tilde z}_- (E,Q^2 \vert \;  Q_1^2,Q_2^2) < z < {\tilde
z}_+(E,Q^2 \vert \; Q_1^2,Q_2^2)
\end{equation}
where
\begin{equation}\label{zpm0}
 {\tilde z}_{\pm}(E,Q^2 \vert \;  Q_1^2,Q_2^2))=
 \frac{1}{2} \left(1+\frac{q_+ q_-}{Q^2}  \pm
 \sqrt{\left(1-\frac{Q^2}{E^2}\right)
      \left(1-\frac{q_{+}^2}{Q^2}\right)
       \left(1-\frac{q_{-}^2}{Q^2}\right)}\right)\, ,
\end{equation}
$q_\pm = \sqrt{Q_1^2} \pm \sqrt{Q_2^2}$. The restriction (\ref{splitint0})
follows from the requirement of positivity of the square of transverse
momentum generated in the decay. Let us also note that, as is immediately
evident from (\ref{zpm0}), one has $Q_1+Q_2\leqslant Q$.

From equations (\ref{splitint0},\ref{zpm0}) it follows that the full
specification of a decay requires, generally speaking, knowledge of invariant
masses
of the offspring partons. In our simulations we employ the so-called 'global'
'constrained' prescription  that uses both the exact restriction
(\ref{splitint0},\ref{zpm0}) and its simplified version in
which $Q^2_{1,2}$ are replaced by the minimal invariant mass $Q_h^2/4$:
\begin{equation}\label{splitint}
z_- (E,Q^2 \vert \;  Q_h^2) < z < z_+(E,Q^2 \vert \; Q_h^2)
\end{equation}
where
\begin{equation}\label{zpm}
 z_{\pm}(E,Q^2 \vert \;  Q_h^2) 
\equiv {\tilde z}_\pm \left ( E,Q^2 \vert \;  \frac{Q_h^2}{4},\frac{Q_h^2}{4} \right)
 =\frac{1}{2} \left(1 \pm \sqrt{\left(1-\frac{Q^2}{E^2}\right) 
 \left(1-\frac{Q_h^2}{Q^2}\right)} \right)
\end{equation}

In particular, the simplified restrictions (\ref{splitint},\ref{zpm}) are used
when calculating the Sudakov formfactor
\begin{equation}\label{sff}
S(Q^2 \vert \; E, Q^2_{\rm max};Q_h^2)= \\ \exp \left[
-\int_{Q^2}^{Q_{\rm max}^2}\frac{dt^2}{t^2} \int_{z_{-}(E,\;t^2
\vert \; Q_h^2)}^{z_{+}(E,\;t^2 \vert \;  Q_h^2)} dz
\frac{\alpha_s[z(1-z)t^2]}{2\pi} N_c K(z) \right].
\end{equation}
The equation (\ref{sff}) defines the probability to split into 
two gluons with virtualities $Q_h/2$ 
for a gluon with the energy $E$ having a virtuality $Q^2$ restricted by condition  
$Q^2_h<Q^2 < Q^2_{\rm max}$. In (\ref{sff}) $N_c=3$ is the number 
of colors and $K(z)=\dfrac{1}{z(1-z)} - 2 + z(1-z)$
is the gluon splitting function.

\medskip

The explicit procedure we employ to generate an in-vacuum mass-ordered gluon
cascade is as follows:

\begin{enumerate}
\item{First one draws, using (\ref{splitint},\ref{zpm},\ref{sff}), the scale
$Q^2$ at which the gluon under consideration branches into two
new gluons. Let us note that at this step one fixes the lifetime of the gluon
$\tau = E(1/Q^2-1/Q^2_{\rm par})$. }
\item{One draws the value of the splitting variable $z$ determining the energies
of the offspring gluons $E_1=zE$ and $z_2=(1-z)E$.}
\item{With the energies of the offspring gluons fixed, one draws their final
invariant masses $Q_{1,2}^2$ and, therefore, fix their lifetimes
$$
\tau_{1,2}=E_{(1,2)}\left( \frac{1}{Q_{(1,2)}^2}-\frac{1}{Q^2}
\right )
$$
The values of $Q_{1,2}^2$ are accepted if they do not violate (\ref{zpm0}).
In other case the maximum of $Q_{1,2}$ is redrawn until the condition
(\ref{zpm0}) is satisfied\footnote{Let us note that such a procedure 
tends to exclude particles that simultaneously have big energies and 
virtualities.
Indeed, let us assume that we have drawn some $Q\approx E$. This 
immediately leads to $z\approx 1/2$ and, therefore, makes it impossible 
to get $E_1\approx E$ with $Q_1\approx E_1$. At the same time if we 
drew $Q_1=Q-Q_h/2$ and $Q_2=Q_h/2$, we would get the case 
from (\ref{zpm}): $z\approx 1-Q_h/E$.}.}
\item{Finally, one ensures angular ordering by drawing the splitting 
variables for the decays of the offspring gluons $z_{1,2}$ and accepting 
them if they satisfy the inequalities
\begin{eqnarray}
z_1(1-z_1) & > & \frac{1-z}{z} \left( \frac{Q_1^2}{Q^2} \right) \nonumber \\
z_2(1-z_2) & > & \frac{z}{1-z} \left( \frac{Q_2^2}{Q^2} \right)
\nonumber
\end{eqnarray}
correspondingly.}
\end{enumerate}

\subsection{Decoherence of in-medium QCD cascades}\label{dc}

We have already mentioned that quantum coherence and resulting angular ordering
of gluon emissions plays a crucial role in the physics of QCD
jets. The violent environment created in ultrarelativistic heavy ion collisions
acts as a source of random energy-momentum and color with respect to
in-vacuum processes. This kind of random impact tends to ruin the phase tuning
lying at the heart of quantum interference phenomena. The situations in which
random external influences destroy the interference-related effects are very
common in solid state physics. A well-studied example is provided by studying
the physics of weak localization in the presence of random external impacts, see
e.g. \cite{B84,CS86}. Thus it is intuitively quite clear that the quantum
coherence effects  should be broken in the violent environment created in heavy
ion collisions. This was first explicitly mentioned in  \cite{JEWEL}, in
which the origin of decoherence and ensuing disappearance of angular ordering
was related to rescattering of cascading particles on scattering centers in the
medium.

It is well known that in plasma physics the mean-field effects are typically
much stronger than those related to scattering. It therefore of interest to
point out to another source of decoherence, the randomness of the color of mean
field. A quantitative argument is provided by comparing the characteristic
times of color and momentum diffusion in dense strongly interacting matter,
$t_c$ and $t_p$ correspondingly, that were calculated for quark-gluon plasma in
\cite{SG}:
\begin{eqnarray}\label{timescales}
t_p & \thickapprox & \left[4 \alpha_s^2 T \ln(1/\alpha_s) \right]^{-1} \nonumber
\\
t_c & \thickapprox & \left[ 3 \alpha_s T \ln(m_E/m_M) \right]^{-1},
\end{eqnarray}
where $m_E$ is an electric screening mass and $m_M$ - a (non-perturbative)
magnetic screening one. Obviously the pace of color randomization is
much faster than that of momentum evolution and, therefore, collisional energy
loss, $t_c \ll t_p$. This situation is similar to decoherence
related to rotation of spin by random external magnetic filed \cite{B82}.
Therefore, when describing a propagation of a test high energy colored
particle through plasma, one can assume that at time scales of order of $t_c$
all coherent effects related to precise color matching are
destroyed. The origin of angular ordering in the in-vacuum QCD cascades is
exactly in such kind of precise color matching, therefore at the
timescales $\tau \gtrsim t_c$ the color coherence is broken. Of course, the
estimates for the timescales in (\ref{timescales}) refer to
equilibrated QCD matter considered in the hard loop approximation and are thus
not too realistic. In the context of the present paper it is only
important that (\ref{timescales}) points out to color randomization as the
fastest source of decoherence in the hot dense fireball.

In what follows we shall use a simplest assumption that angular ordering is
always broken for gluon decays in QCD cascades that take place inside
the hot dense medium. In this paragraph we consider the situation in which such
breaking of angular coherence is the only medium-related effect in
the gluon cascade. It is known for a long time that in the absence of angular
ordering the properties of QCD cascades are very different from
those when angular ordering is imposed \cite{BS87}. In particular, the jet
multiplicities decrease, but the strongest effect is related to the
changes in intrajet energy distribution which experiences dramatic softening
\cite{BS87}. As in our considerations the effects of breaking of
color ordering are present only for decays taking place inside the medium, the
cumulative effect on energy distribution of final prehadrons, the
strength of the above-mentioned softening should depend on the relative
proportion of vertices generated inside the medium or, in other terms, on
the distance $L$ from the cascade origin to the border of the fireball.

\begin{figure}[h]
 \centering
 \includegraphics[width=\textwidth]{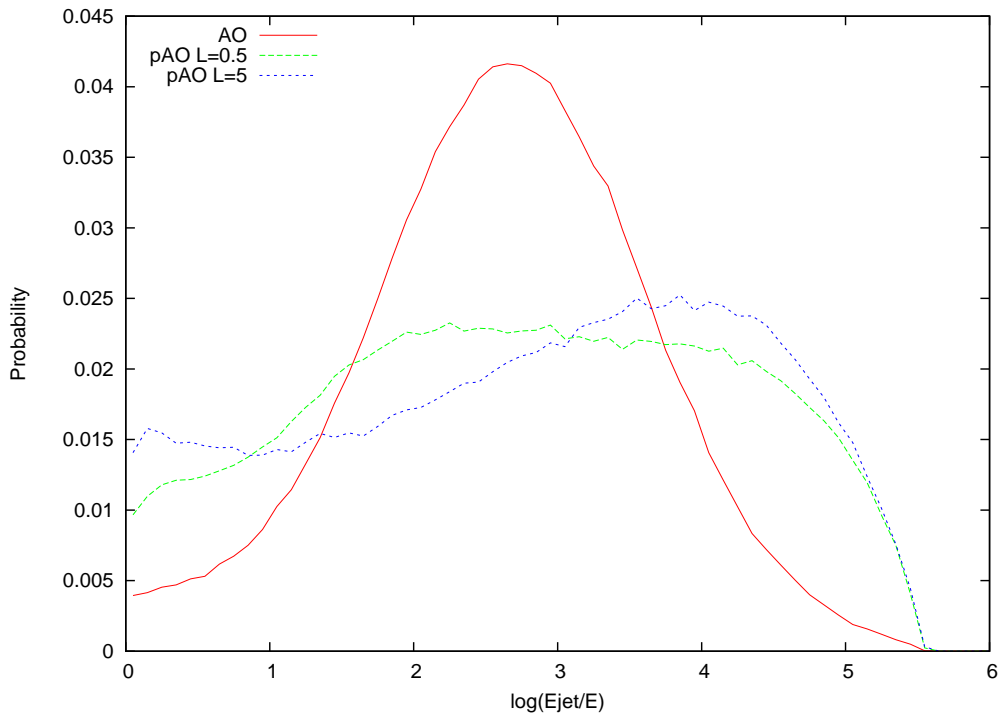}
 \caption{Rapidity distribution $P(y)$ of final prehadrons: 
1) $L=0\,{\rm fm}$, full angular ordering, red, solid; 2) $L=0.5\,{\rm fm}$,
 partial angular ordering, green, dashed; 3) $L=5\,{\rm fm}$, partial angular ordering, blue, dotted.}
\label{paonl}
\end{figure}

In Fig.~\ref{paonl} we plot the distributions of final prehadrons in 
the rapidity $y=\ln(E_0/E)$ for gluon jets with the initial energy 
$E_0=100$ GeV and $L=$ 0.5, 1 and 5 fm.

From the distributions in Fig.~\ref{paonl} we see that the effect of 
color decoherence is indeed very strong. Already at $L=$ 0.5 fm 
the distribution is significantly different from the vacuum one and 
at larger $L$ practically saturates. Thus, even in the absence of 
energy loss in terms of energy-momentum, the intrajet properties 
are significantly affected by the medium through color decoherence.

\begin{figure}[h] 
 \centering
 \includegraphics[width=\textwidth]{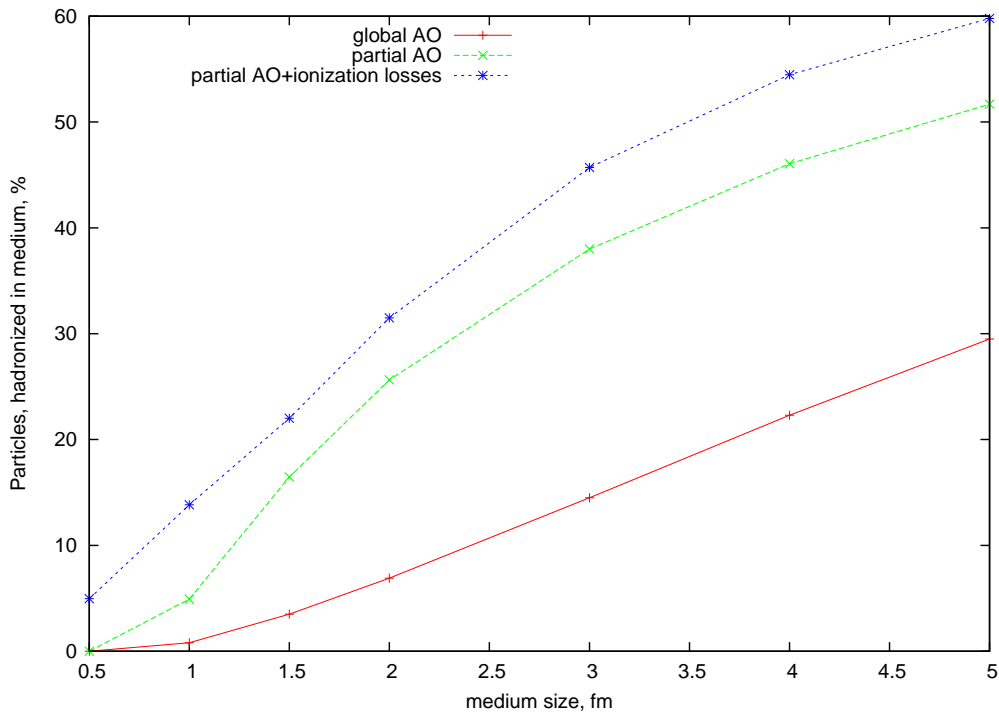}
 \caption{Relative yield of prehadrons formed inside the medium: 1) global angular ordering, red, solid; 2) partial angular ordering, green, dashed;
 3) partial angular ordering and collisional losses, blue, dotted. }
 \label{preformin}
\end{figure}

The changes in distribution over the rapidity $\ln (E_{\rm jet}/E)$ 
shown in Fig.~\ref{paonl} reflect important changes in 
the spatiotemporal pattern of the jet. Indeed, from the inside-outside 
nature of particle production in which soft particles are produced 
faster than hard ones, cf. Eq.~(\ref{formtime}) one can conclude that 
the substantial softening of the rapidity distribution shown in 
Fig.~\ref{paonl} reflects an increased yield of
short formation times of prehadrons. Thus it is of interest to consider 
the yield of prehadrons formed inside the medium of given length. This
yield is plotted in Fig.~\ref{preformin} for jets with and 
without angular ordering.

\subsection{Collisional losses in the in-medium QCD cascades}\label{ic}

In this paragraph we shall make a crude estimate of the efects of the part of
medium-induced energy loss which is proportional to the distance $l$
that a parton or prehadron travel inside the medium. For simplicity, in analogy
with QED cascades in matter, we shall term these losses as
collisional losses. Radiative medium-induced losses are, at least at small times,
proportional to $l^2$ and require a separate treatment. Usually
one considers the radiative losses as dominant, so our restriction to the collisional
ones provides an estimate of the minimal medium-induced energy
loss.

In analogy with QED showers \cite{L44,V57}, the partonic intrajet energy losses
$\Delta E_p$ are described by the continuous distribution 
${\cal P}(\Delta E_p,l \vert \; \mu_{\Delta E}, \sigma_{\Delta E})$. 
The parameters $(\mu_{\Delta E}, \sigma_{\Delta E})$ describe the mean energy 
loss and the corresponding standard deviation at the distance of 1 fm.
In our simulations we use a simple Gaussian parametrization of 
${\cal P}(\Delta E,l \vert \; \mu_{\Delta E}, \sigma_{\Delta E})$:
\begin{equation}
 {\cal P}(\Delta E_p,l \vert \; \mu_{\Delta E}, \sigma_{\Delta E}) 
 = \nu \delta(\Delta E_p) + \frac{1}{\sqrt{2 \pi} \sigma_{\Delta E} l}
 \exp \left \{ -\frac{(\Delta E_p - \mu_{\Delta E} l)^2}{2 \sigma^2_{\Delta E} l^2}  \right \}
\Theta(\Delta E_p)
\end{equation}
where
$$
\nu = \frac{1}{\sqrt{2 \pi}} \int^\infty_{1} dx \exp(-x^2/2)
=\frac{1}{2}(1- {\rm erf}(\sqrt{2}/2))\approx 0.159
$$
as we use only the case $ \mu_{\Delta E}=\sigma_{\Delta E}$.
The numerical values we chose in most of the simulations were 
$\mu_{\Delta E}=1\;{\rm GeV}/{\rm fm}$ and 
$\sigma^2_{\Delta E} = 1 \; {\rm GeV}^2/{\rm fm}$.

The losses $\Delta E_{h}$ experienced by prehadrons formed inside the medium on
their way out of the hot fireball were simply taken to be
proportional to the distance they cover inside the medium, $\Delta E_h = 1\;
{\rm GeV}/{\rm fm} \cdot l$.

\begin{figure}
\centering
\includegraphics[width=\textwidth]{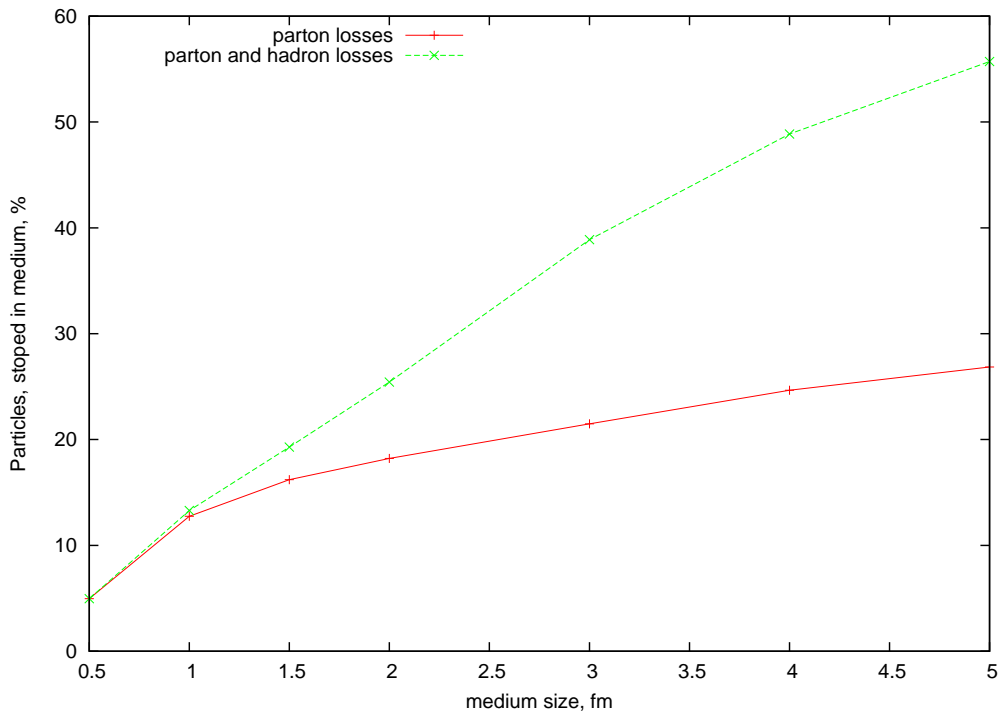}
\caption{Relative yield of particles stopped inside the medium: 1) parton losses, red, solid; 2) parton and hadron losses, green, dashed.}
\label{StopCountL}
\end{figure}

Taking into account the softness of the majority of produced prehadrons 
that takes place even without taking into account the energy losses, just
because of decoherence, an interesting question to address is how many 
prehadrons have enough energy to leave the medium when the energy losses
are switched on. The simplest crude criterion of stopping one can use is 
$E_{\rm crit} = Q_h/2$ so that the momentum of the corresponding parton
or prehadron is zero. The corresponding yields are shown 
in Fig.~\ref{StopCountL}.

\begin{figure}
\centering
\includegraphics[width=\textwidth]{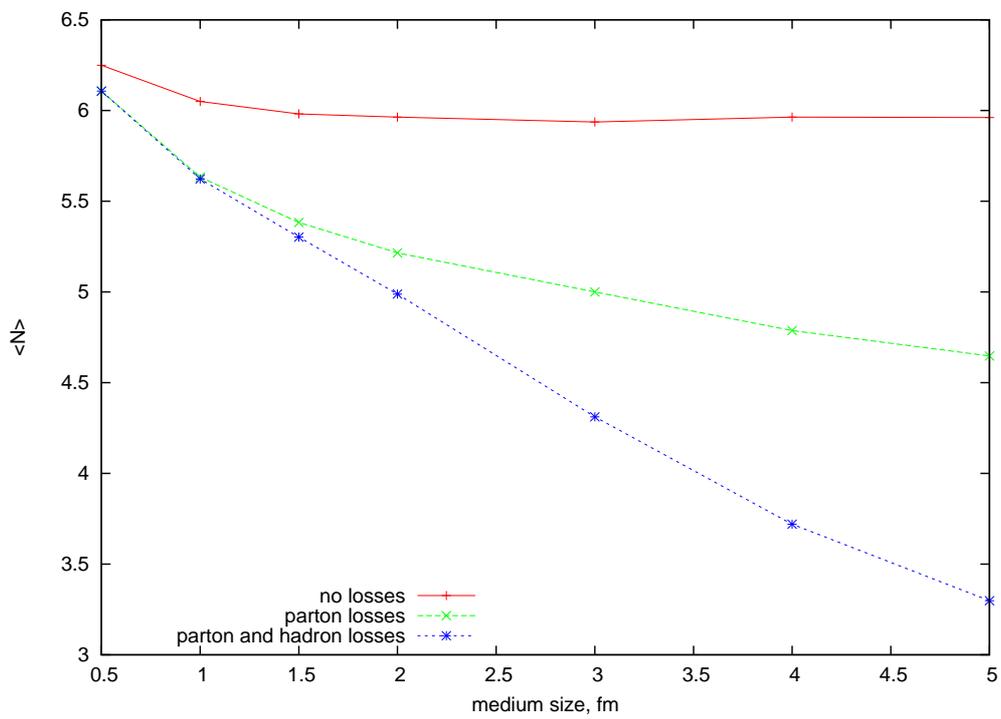}
\caption{Mean jet multiplicity as a function of medium size: 1) 
no losses, red, solid; 2) intracascade parton losses, green, dashed; 
3) intracascade and final state losses, blue, dotted.}
\label{FigMultL}
\end{figure}

We see that at large enough distances to the border of the fireball 
the effect is very significant: at $L=5 \; {\rm fm}$ more than half 
of the final prehadrons are stopped inside the fireball. The effect 
of stopping is also clearly seen in Fig.~\ref{FigMultL} in which 
we plot the average multiplicity
of final prehadrons inside the cone $\theta_{\rm jet}=0.7$.

\begin{figure}
\centering
\includegraphics[width=\textwidth]{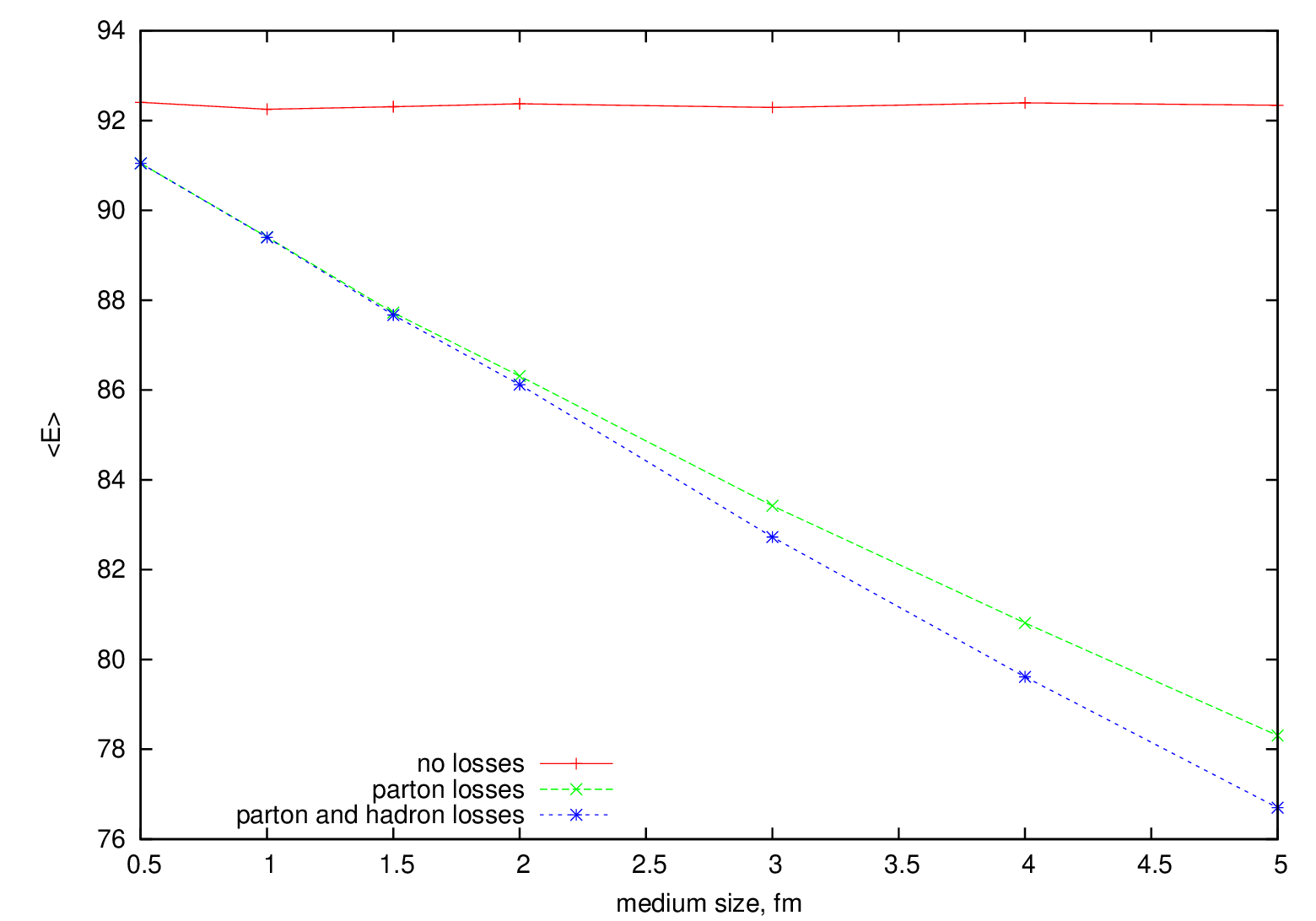}
\caption{Mean jet energy as a function of medium size: 1) no losses, red, solid,; 
2) intracascade parton losses, green, dashed; intracascade and final state losses, blue, dotted.}
\label{FigEconeL}
\end{figure}

For completeness in Fig.~\ref{FigEconeL} we show the dependence of average energy within 
the cone of $\theta_{\rm jet}=0.7$ on $L$.

\begin{figure}
\centering
\includegraphics[width=\textwidth]{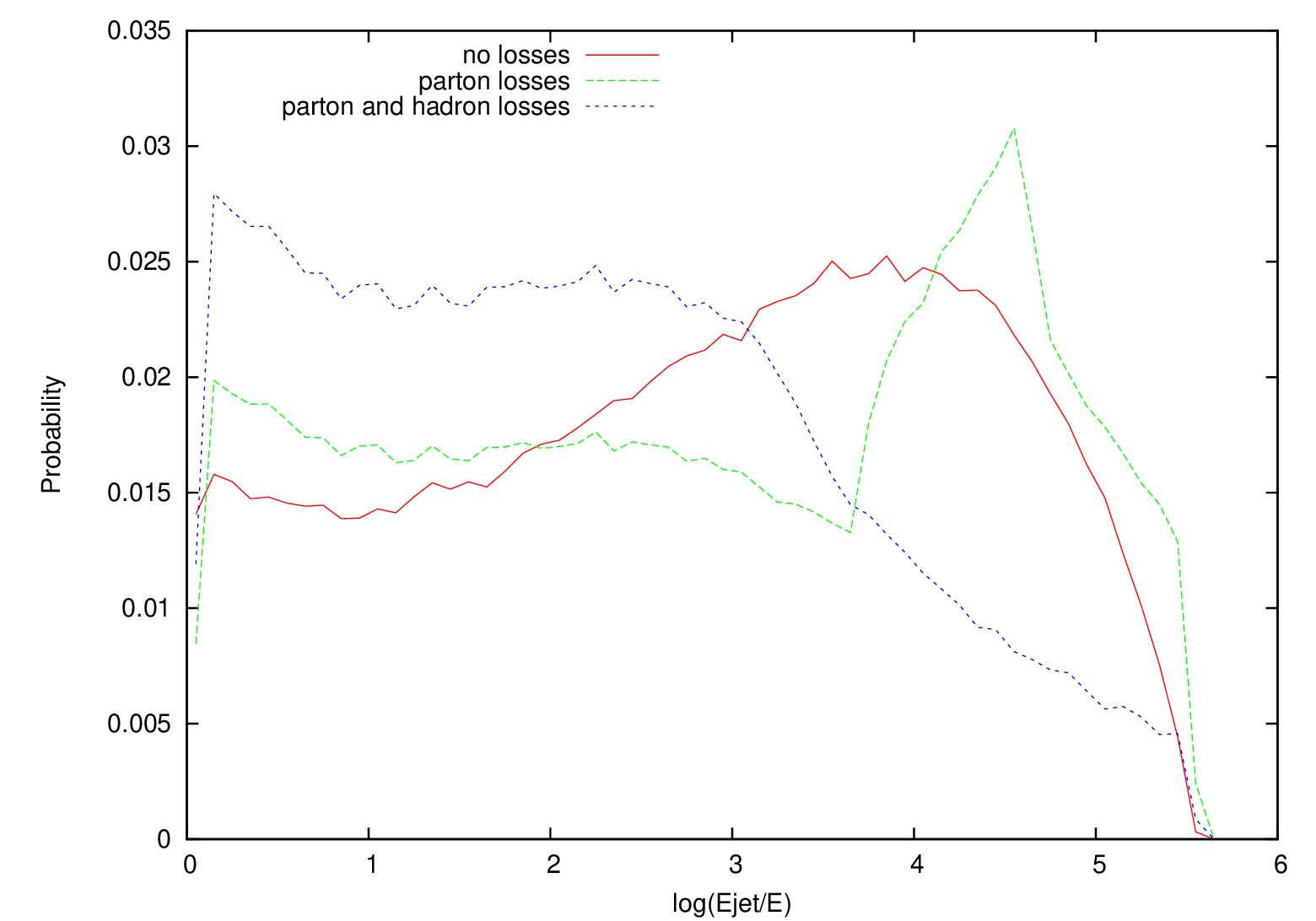}
\caption{Rapidity distribution $P(y)$ of final prehadrons, cascade with
 partial angular ordering: 1) no losses, red, solid;  2) intracascade parton losses, green, dashed; 
 3) intracascade and final state losses, blue, dashed.}
\label{PlnE}
\end{figure}

\begin{figure}
\centering
\includegraphics[width=\textwidth]{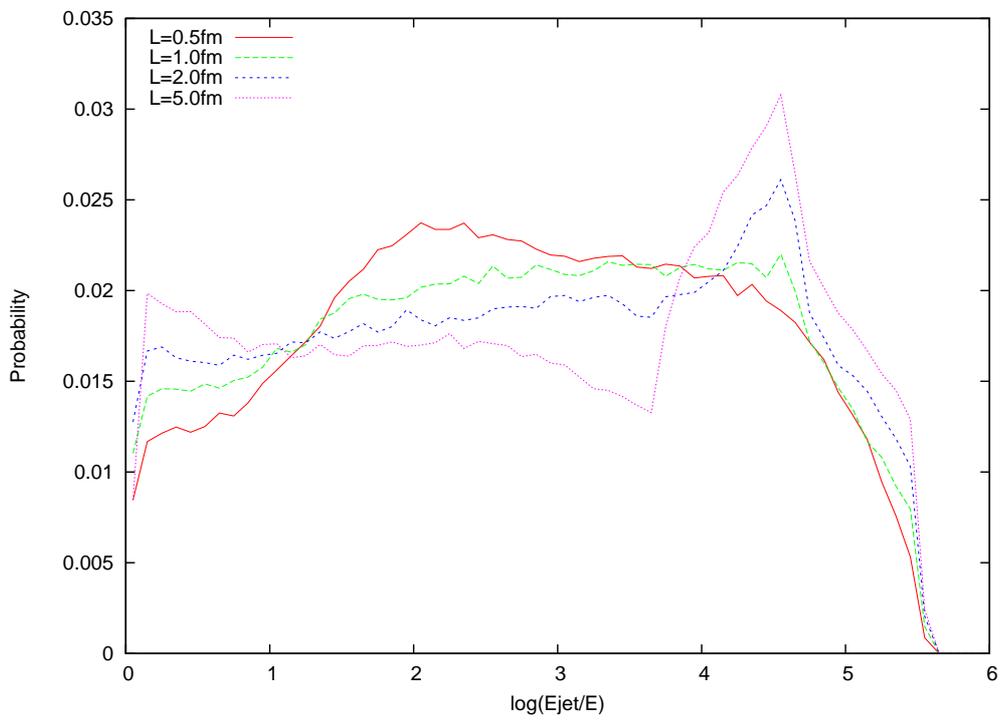}
\caption{Rapidity distributions at different medium length $L$}
\label{PlnE_L}
\end{figure}

The additional softening incurred by energy losses does considerably 
change the shape of the rapidity distributions of final prehadrons. In Fig.~\ref{PlnE} 
we plot the corresponding distributions at $L=5\;{\rm fm}$ and 
in Fig.~\ref{PlnE_L} we illustrate the dependence of rapidity 
distribution taking into account all sources
of energy loss on $L$.

\section{Conclusions}

The main goal of the present study was to analyze some medium-induced changes 
in the the evolution of the in-medium QCD cascade. 

In the focus of our attention was a detailed analysis of the medium-induced color decoherence and the ensuing
disruption of angular ordering for the decays taking place inside the medium. 
We have shown that this effect leads to the substantial softening of the intrajet 
rapidity spectrum. 

Another quantitatively important effect considered in the paper is that of the 
collisional losses of cascade gluons and of those final prehadrons that happened to be formed inside 
the medium. We have shown that, at variance with the widespread opinion that the dominant majority 
of the final hadrons form outside the hot and dense fireball, the yield of prehadrons formed inside the 
fireball is quite significant. Combined with the effect of partonic collisional loss, this leads to the
substantial deformation of the rapidity spectra.

To make the analysis presented in this paper more complete one has, of course, take into account the
medium-induced radiative losses. Work on this problem is currently in progress.

\begin{center}
{\bf Acknowledgements}
\end{center}

We are grateful to I.M. Dremin for very useful discussions.

The work was supported by the RFBR grants 06-02-17051, 08=02-91000, 09-02-00741
and the RAS program "Physics at the LHC collider".

\renewcommand{\theequation}{A.\arabic{equation}}
\setcounter{equation}{0}
\section*{Appendix}
Let us discuss the approximations which made when ``angular orderung''
conditions is introduced. Recall that the ``angular'' varible\footnote{This
is exactly the same variable for which the ordering was originally derived in
\cite{MullerAO}} which is used in Herwig and similar is written as

\begin{equation}
 \xi \equiv \frac{(p_b,p_c)}{E_b E_c}=1-\frac{(\vec{p}_b,\vec{p}_c)}{E_b E_c}=
1-\frac{1}{2}\frac{\vec{p}_a^2-\vec{p}_b^2-\vec{p}_c^2}{E_b E_c}, \label{xi}
\end{equation}

\noindent where we have used

\begin{equation}
 \vec{p}^2_a=(\vec{p}_b + \vec{p}_c)^2 = \vec{p}^2_b + \vec{p}^2_c +2(\vec{p}_b,\vec{p}_c)
\label{scalarProd}
\end{equation}

\noindent The initial upper limit for $\xi$ usually supposed as 1. This means
that we consider the openning angles $\theta_a$ to be less then
$\pi/2$ only. In this area the conditions of angular ordering ($\theta$) and
$\sin \theta$ (or $p_t$ or $\sin^2 \theta$) are equivalent. In terms
of momenta we have for $\sin^2 \theta$

\begin{equation}
\sin^2 \theta_a=1-\frac{(\vec{p}_b,\vec{p}_c)^2}{\vec{p}^2_b\vec{p}^2_c}=
1-\frac{1}{4}\frac{(\vec{p}_a^2-\vec{p}_b^2-\vec{p}_c^2)^2}{\vec{p}_b^2\vec{p}_c^2}
\end{equation}

\noindent where we again used (\ref{scalarProd}).

\begin{figure}[h]
 \centering
 \includegraphics[width=0.5\textwidth]{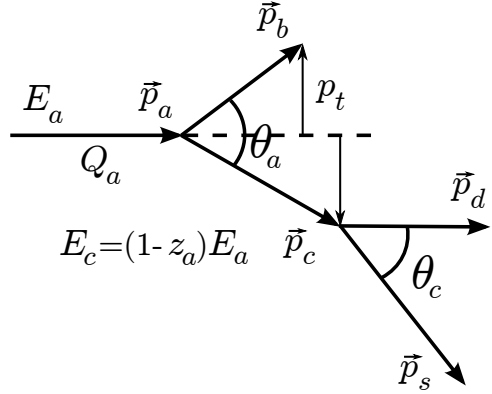}
 \caption{Angular ordering}
\end{figure}

As for considered angles  $\vec{p}_a^2-\vec{p}_b^2-\vec{p}_c^2>0$ we can rewrite
the ordering condition as following (see Figure for parton indexes)

\begin{equation}
\frac{\vec{p}_c^2-\vec{p}_s^2-\vec{p}_d^2}{|\vec{p}_s||\vec{p}_d|} \geqslant
\frac{\vec{p}_a^2-\vec{p}_b^2-\vec{p}_c^2}{|\vec{p}_b||\vec{p}_c|}
\label{exactAO}
\end{equation}
\noindent and for $\xi$ variable

\begin{equation}
\frac{\vec{p}_c^2-\vec{p}_s^2-\vec{p}_d^2}{E_s E_d} \geqslant
\frac{\vec{p}_a^2-\vec{p}_b^2-\vec{p}_c^2}{E_b E_c} \label{WebberAO1}
\end{equation}

\noindent Comparing (\ref{exactAO}) and (\ref{WebberAO1}) we see that
$\xi$-ordering coinside with true angular ordering if we suppose $E\approx
|\vec{p}|$ in denominators of (\ref{exactAO}). Which condition is stronger
depends on the four ratios $Q/E$ and can be different for different
cases.

From other side
\begin{equation}
\vec{p}_a^2-\vec{p}_b^2-\vec{p}_c^2 =
 E_a^2-E_b^2-E_c^2-Q_a^2+Q_b^2+Q_c^2=
 2E_b E_c-Q_a^2+Q_b^2+Q_c^2
\end{equation}
where we have used energy conservation $E_a=E_b+E_c$. And we can rewrite
condition (\ref{WebberAO1}) as following

\begin{equation}
\frac{Q_c^2-Q_d^2-Q_s^2}{E_s E_d} \leqslant \frac{Q_a^2-Q_b^2-Q_c^2}{E_b E_c}
\label{WebberAO2}
\end{equation}
or taking into account that $E_s E_d=z_c(1-z_c)E^2_c$ and $E_b=z_a E_a=z_s
\dfrac{E_c}{(1-z_a)}$,

\begin{equation}
\frac{Q_c^2-Q_d^2-Q_s^2}{z_c(1-z_c)} \leqslant
\frac{(Q_a^2-Q_b^2-Q_c^2)(1-z_a)}{z_a} \label{WebberAO3}
\end{equation}

\noindent If we now suppose that $Q_a^2 \gg Q_b^2 + Q_c^2$ and $Q_s^2 \gg Q_d^2
+ Q_a^2$ we get the conditions which is used in ``Virtuality
ordered`` Pythia as angular ordering condition
\begin{equation}
\frac{Q_c^2}{z_c(1-z_c)} \leqslant \frac{Q_a^2(1-z_a)}{z_a} \label{PythiaAO}
\end{equation}


\begin{thebibliography}{99}

\bibitem{E09}
d'Enterria D arXiv:0902.2011 [nucl-ex]

\bibitem{BSZ00}
R. Baier, D. Schiff, B.G. Zakharov, {\it Ann. Rev. Nucl. Part. Sci.}\; {\bf 50}
(2000), 37

\bibitem{KW03}
Kovner A, Wiedemann U A, "Gluon radiation and parton energy loss",
in R.C. Hwa and X.-N. Wang (eds) "Quark Gluon Plasma 3", 192 (2003),
arXiv:hep-ph/0304151

\bibitem{GVWZ03}
M. Gyulassy, I. Vitev, X.-N. Wang, B-W. Zhang, "Jet quenching and radiative
energy loss in dense nuclear matter", 
in R.C. Hwa and X.-N. Wang (eds) "Quark Gluon Plasma 3", 123 (2003),
arXiv:nucl-th/0302077

\bibitem{SS07}
Casalderrey-Solana J, Salgado C A, arXiv:0712.3443 [hep-ph]

\bibitem{W09}
Wiedemann U A arXiv:0908.2306 [hep-ph]

\bibitem{M10}
Majumder A arXiv:1002.2206 [hep-ph]


\bibitem{DKMT91}
Yu.L. Dokshitzer, V.A. Khoze, A.H. Mueller and S.I. Troyan, {\it Basics of
perturbative QCD}, Edition Frontiers, 1991

\bibitem{JSTAR}
J. Putschke, {\it Eur. Phys. J.}\, {\bf C61} (2009), 629\\
E. Bruna, {\it Jet fragmentation in STAR going from p+p to Au=Au},
arXiv:0905.4763[nucl-ex]\\
E. Bruna, {\it Measurements of jet structure and fragmentation from full jet
reconstruction in heavy ion collisions at RHIC},
arXiv:0907.4788[nucl-ex]

\bibitem{JLHC}
M.L. Noriega, {\it Eur. Phys. J.}\, {\bf C49} (2007), 315 \\
J.L. Klay, {\it Nucl. Phys.}\, {\bf A787} (2007), 52


\bibitem{PYTHIA}
Sjostrand T, Mrenna S, Skands P {\it JHEP}\ 0605:026 (2006) 

\bibitem{HERWIG}
Marchesini G, Webber B R, Abbiendi G, Knowles I G, Seymour M H, Stanco L {\it
Comput. Phys. Commun.}\; {\bf 67} 465 (1992); 
Corcella G, Knowles I G, Marchesini G, Moretti S, Odagiri K, Richardson P,
Seymour M H, Webber B {\it JHEP}\; {\bf 0101} 010 (2001); 
Corcella G, Knowles I G, Marchesini G, Moretti S, Odagiri K, Richardson P,
Seymour M H, Webber B hep-ph/0210213


\bibitem{PYQUEN}
I.P. Lokhtin, A.M. Snigirev {\it Eur. Phys. J.}\; {\bf C45} (2006), 211

\bibitem{JEWEL}
Zapp K, Ingelman G, Rathsman J, Stachel J, Wiedemann U A, {\it Eur. Phys. J}
{\bf C60} 617 (2009)

\bibitem{QPYTHIA}
Armesto N, Cunqueiro L, Salgado C A, {\it Eur. Phys. J} {\bf C63} 679 (2009)

\bibitem{QHERWIG}
Armesto N, Corcella G, Cunqueiro L, Salgado C A, {\it JHEP} 0911:122 (2009)


\bibitem{coscad}
S.Z. Belenky, {\it Shower Processes in Cosmic Rays}, Moscow, Gostechizdat,
1948 (in Russian)

\bibitem{pionizationNAO}
Dremin I M {\it JETP Lett.}\, {\bf 31} 185 (1980);
Dremin I M, Leonidov A V {\it Sov. Phys. Usp.}\, {\bf 23} 515 (1980);
Dremin I M, Leonidov A V {\it Sov. Journ. Nucl. Phys.}\, {\bf 35} 247 (1982);
Dremin I M, Leonidov A V {\it Sov. Phys. Usp.}\, {\bf 38} 723 (1995)

\bibitem{pionizationAO}
Leonidov A V, Ostrovsky D M  {\it Phys. Atom. Nucl.}\, {\bf 60} 110 (1997);
Leonidov A V, Ostrovsky D M, {\it Phys. Atom. Nucl.}\, {\bf 62} 701 (1999)


\bibitem{ZCR}
B.G. Zakharov, {\it JETP Lett.}\; {\bf 86} (2007), 444; ibid., {\bf 88} (2008),
781


\bibitem{SG}
A.V. Selikhov, M. Gyulassy, {\it Phys. Lett.}\ {\bf B316} (1993), 373\\
A.V. Selikhov, M. Gyulassy, {\it Phys. Rev.}\ {\bf C49} (1994), 1726

\bibitem{B84}
G. Bergmann, {\it Phys. Rep.}\; {\bf 107} (1984), 1

\bibitem{CS86}
S. Chakravarty, A. Schmid, {\it Phys. Rep.}\; {\bf 140} (1986), 195

\bibitem{B82}
G. Bergmann, {\it Solid State Commun.}\; {\bf 42} (1982), 815

\bibitem{BS87}
M. Bengtsson, T. Sjostrand, {\it Nucl.Phys.}\, {\bf B289} (1987), 810

\bibitem{L44}
L.D. Landau, {\it Journal of Physics USSR}\, {\bf 8} (1944), 204

\bibitem{V57}
P.V. Vavilov, {\it JETP}\, {\bf 32} (1957), 920


\bibitem{MullerAO} 
A.H.~Muller,{\it Phys. Lett.} {\bf B104} (1981) 161


\end{thebibliography}
\end{document}